\documentclass[graybox]{svmult}

\usepackage{mathptmx}       % selects Times Roman as basic font
\usepackage{helvet}         % selects Helvetica as sans-serif font
\usepackage{courier}        % selects Courier as typewriter font
\usepackage{type1cm}        % activate if the above 3 fonts are
\usepackage{makeidx}         % allows index generation
\usepackage{graphicx}        % standard LaTeX graphics tool
\usepackage{multicol}        % used for the two-column index
  \usepackage{natbib}
\usepackage[bottom]{footmisc}% places footnotes at page bottom

\makeindex             % used for the subject index
\begin{document}

\title*{Observed properties of boxy/peanut/barlens bulges}
% Use \titlerunning{Short Title} for an abbreviated version of
% your contribution title if the original one is too long
\author{Eija Laurikainen and Heikki Salo}
% Use \authorrunning{Short Title} for an abbreviated version of
% your contribution title if the original one is too long
\institute{Eija Laurikainen \at Astronomy and Space physics, University of Oulu, FI-90014 Oulu, Finland, \email{eija.laurikainen@oulu.fi}
\and Heikki Salo \at Astronomy and Space physics, Oulu, Finland \email{heikki.salo@oulu.fi}}
%
% Use the package "url.sty" to avoid
% problems with special characters
% used in your e-mail or web address
%
\maketitle

%\abstract*{Each chapter should be preceded by an abstract (10--15.}

\abstract{ We review the observed morphological, photometric, and
  kinematic properties of boxy/peanut (B/P) shape bulges. Nearly half
  of the bulges in the nearby edge-on galaxies have these
  characteristics, which fraction is similar to the observed bar
  fraction in Hubble types earlier than Scd. B/P bulges are generally
  detected in the edge-on view, but it has been recently demonstrated
  that barlenses, which are lens-like structures embedded in bars, are
  the more face-on counterparts of the B/P bulges.  Multi-component
  structural decompositions have shown that B/P/barlens structures are
  likely to account for most of the bulge light, including the
  early-type disks harboring most of the bulge mass in galaxies.
  These structures appear in bright galaxies, in a mass range near to
  the Milky Way mass. Also the other properties of these bulges,
  including morphology (X-shaped), kinematics (cylindrical rotation),
  or stellar populations (old), are similar to those observed in the
  Milky Way. Cool central disks are often embedded in the B/P/barlens
  bulges. Barred galaxies contain also dynamically hot classical
  bulges, but it is not yet clear to what extent they are really
  dynamically distinct structure components, and to what extent stars
  wrapped into the central regions of the galaxies during the
  formation and evolution of bars.  If most of the bulge mass in the
  Milky Way mass galaxies in the nearby universe indeed resides
  in the B/P-shape bulges, and not in the classical bulges, that idea needs
  to be integrated into the paradigm of galaxy formation.  }
%LISAA: about stellar populations

\section{Introduction}
\label{sec:1}

%Barentine & Kormendy (2012, ApJ, 754, 140): call boxy-structures
%as ``boxy pseudobulges'', since it is really a part of the disk

Galaxies in the nearby universe have complex morphological structures
and indeed the concept of the bulge depends strongly on how we define
it. Bulges can be considered simply as an excess flux above the disk,
or they can be defined by detailed morphological, photometric, or
kinematic properties.  An important question is to what extent these
central mass concentrations are associated to the early formative
processes of galaxies, and how much are they modified via internal
dynamical processes. Using the notation by \citet{atha2005} bulges can
be divided to {\it classical bulges, disky pseudobulges, and
  boxy/peanut (B/P) bulges}.  Classical bulges are the most obvious
imprints of galaxy formation at high redshifts, formed in some violent
processes, followed by strong relaxation.  They are structures
supported by velocity dispersion, and also have centrally peaked
surface brightness profiles with high S\'ersic indexes. Pseudobulges
are defined as structures formed by secular evolutionary processes out
of the disk material. The concept of a pseudobulge was introduced by
\citet{kormendy1982,kormendy1983} who defined them as flat structures
in the central parts of the disks, having excess of light above the disk in
the surface brightness profile.  Pseudobulges were later extended to
include also the vertically thick boxy/peanut bulges, generally
associated to bars (Athanassoula 2005; many B/Ps show also X-shape
morphology).  Since then the flat central mass concentrations have
been referred as 'disky pseudobulges'.  A division of pseudobulges to
these two categories is important, because their observed properties
are very different.

The concept of a pseudobulge was created having in mind the relaxed
universe where slow secular evolutionary processes are prevalent, not
the early gas rich clumpy universe, or the universe where galaxy
mergers dominate the evolution. However, it appears that similar
internal mechanisms which take place in the local universe, like bar
instabilities, can occur also at high redshifts, but in a much more
rapid dynamical timescale.  As discussed by Brooks $\&$ Christensen
and by Bournaud in Section 6, at high redshifts also star formation
and the different feedback mechanisms are faster and more
efficient. All this can lead to the formation of pseudobulges even at
high redshifts, for which reason associating the different bulge types
to a unique formative process of bulge is not straightforward. In
spite of that the above definition of bulges may still be a good
working hypothesis at all redshifts, and give useful insight to the
theoretical models.

This article reviews the observational properties of B/P bulges and
also gives a historical perspective for the discovery of the
phenomenon. The theoretical background of the orbital families is
given by Athanassoula in Section 6.3. We have the following questions
in mind: (a) what is the observational evidence that B/P bulges are
vertically thick inner parts of bars and what are the relative masses
of the thin and thick bar components?  (b) Are the B/P structures the
only bulges in galaxies with this characteristic, or do the same
galaxies have also small classical bulges embedded in the B/P bulges?
(c) What is our understanding of this phenomenon both in the edge-on
and in face-on views? Although the topic is B/P bulges, connections
are made also to other type of bulges if that is needed for
understanding the phenomenon.  Examples of boxy/peanut and X-shape
bulges are shown in Figure 1. The unsharp masks further illustrate the
X-shape, which may appear weakly also in some bulges with apparent
boxy appearance.

An outstanding recent discovery of the Milky Way (discussed in Section
4) is that it harbors a boxy \citep{dwek1995} or even an X-shape bulge
\citep{shen2012,wegg2013}, covering most of the central mass
concentration in our Galaxy. In fact, the model by \citet{shen2010}
which explains most of the observed properties of the Milky Way, does
not have any classical bulge.  We will discuss the observations which
suggests that actually a large majority of the nearby galaxies in the
Milky Way mass range might contain similar boxy or peanut shape
bulges, where most of the bulge mass resides.

Good previous reviews of B/P bulges are those by \citet{combes1981}, \citet{atha2005},
and \citet{bebattista2006}.  Critical points in
the interpretation of B/P bulges are given in the review by \citet{graham2011}.

%--------------------FIG 1 examples of B/P bars
\begin{figure}[t]
%\sidecaption[t]
% Use the relevant command for your figure-insertion program
% to insert the figure file.
% For example, with the option graphics use
\includegraphics[scale=.60]{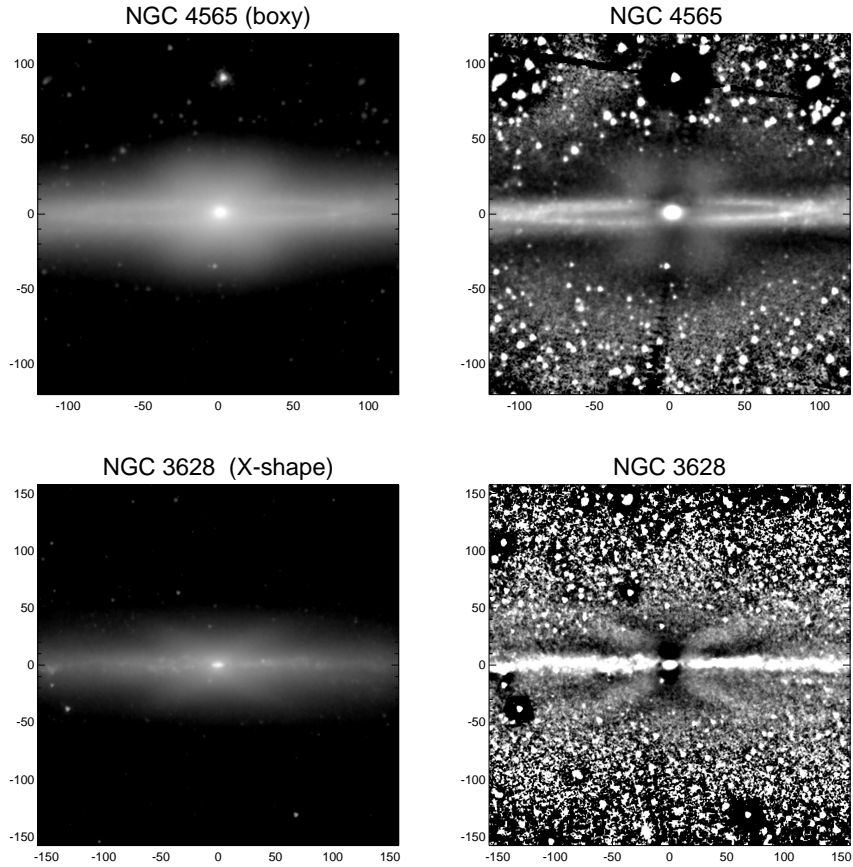}
%
% If no graphics program available, insert a blank space i.e. use
%\picplace{5cm}{2cm} % Give the correct figure height and width in cm
%
%\caption{Please write your figure caption here}
\caption{Examples of boxy (NGC 4565) and peanut (NGC 3628) shape bulges,
using the 3.6$\mu$m Spitzer space telescope images. Also shown are the unsharp mask images of
the same galaxies, which in both galaxies show an X-shape morphology. The units
of the x- and y-axis are in arcseconds.
}
\label{fig:1}       % Give a unique label
\end{figure}

%---------------------------
%\vspace{0.3cm} KUVA$_1$: two edge-on galaxies, NGC 4565 with a boxy
%bulge, and NGC 3628 with an X-shape bulge. Unsharp mask images are
%also shown, filtering out the sharp features of the galaxies. The
%images are 3.6$\mu$m images taken from the Spitzer Survey of Stellar
%Structure in galaxies (S$^4$G, Sheth et al. 2010)
\vspace{0.3cm}

\section{Discovery of B/P bulges}
\label{sec:2}

The earliest notion of boxy bulges goes as far as to 1959, when the
Burbidges recognized such a structure in NGC 128, which galaxy was
later shown in the Hubble Atlas of Galaxies by
\citet{sandage1961}. Some years later \citet{devauc1974} paid
attention to similar structures in some edge-on galaxies, at a time
when bulges were generally thought to be like small ellipticals
sitting in the middle of the disk. An interesting new explanation for
the B/P bulges was given by Combes $\&$ Sanders in 1981: using N-body
simulations they showed that when stellar bars form in galactic disks,
in the edge-on view they reveal boxy or peanut shapes, similar to
those seen in real galaxies. Soon after that a competing
astrophysically interesting explanation for the formation of B/Ps was
given by \citet{binney1985}, who suggested that they might be formed
by violent or soft merging of satellite galaxies. Support for the
scenario by Combes $\&$ Sanders came from the observation that NGC
4565 and NGC 128, with obvious boxy bulges in morphology, appeared to
have cylindrical rotation
\citep{kormendyilling1982,kormendyilling1983}, which means that the
rotational velocity depends only little on the vertical height from
the equatorial plane. Kormendy $\&$ Illingworth argued that
cylindrical rotation might actually be a typical characteristic of
boxy bulges, but not of elliptical galaxies.  It is also worth
mentioning that \citet{bertola1977} had already shown that the bulge
in NGC 128 is fast rotating and therefore cannot be dynamically
hot. However, it is worth noticing that not all bars seen end-on are
perfectly round prolate structures.

After the first discoveries of B/P bulges in individual galaxies,
systematic studies of B/Ps in galaxies seen in the edge-on view were
carried out by \citet{jarvis1986} and \citet{shaw1990}.
They suggested that B/Ps are concentrated to early-type disk
galaxies (S0-Sab), and that peanuts are more common than boxy
bulges, particularly in the late-type galaxies. They also showed
evidence that galaxy environment (cluster/non-cluster, number of
nearby companions) is not critical for the formation of the B/P
bulges. These observations supported the idea that the B/P-structures
indeed form part of the bar as suggested by \citet{combes1981}.
However, this interpretation was not generally accepted at
that time, partly because also $20\%$ of the elliptical galaxies
appeared to be boxy \citep{lauer1985}. \citet{bender1989} even
speculated that boxy ellipticals might have a similar origin as the
cylindrically rotating bulges in disk galaxies. Anyway, as the
fraction of galaxies in which B/Ps were identified was only $\sim$
20$\%$, their exact interpretation would not significantly alter the
estimated total mass fraction of classical bulges in the nearby
universe.

This picture was changed by the studies of \citet{dettmar1988}
and \citet{lutticke2000a}, based on more
complete galaxy samples.  Inspecting the isophotal contours of the
galaxies they found that even 45$\%$ of all S0-Sd galaxies have B/Ps,
which is already close to the fraction of barred galaxies: the
somewhat larger number of detected bars ($\sim$50-60$\%$) could be
easily understood by the aspect angle of the bar. Namely, bars seen
end-on view (along the bar major axis) would look like round
structures, similar to the classical bulges (see the simulation model
by Athanassoula 2005 in Fig. 2).

\citet{lutticke2004} discovered also a new category
of B/P bulges, the so called 'thick boxy bulges'. Such bulges (see
Fig. 3) are thought to be too extended to form part of the bar, and
they also show asymmetries, or signs of recent mergers, which all
means that they are not dynamically settled. Although they form only
a minority of all B/Ps, they indicate that B/P bulges are not a
uniform group of structures. In fact, in the prototype galaxy of
'thick boxy bulges', NGC 1055 \citep{shaw1993}, the bulge looks very much
like a thick disk. This boxy bulge is rotating cylindrically, even in
those regions of the observed light distribution where the bulge
appears neither boxy nor peanut. Such rotation would be natural even
if that structure were interpreted as a thick disk. This
interpretation would be interesting, because in that case in
traditional terms that galaxy barely has a bulge. In some galaxies 
even X-shape is visible in the 'thick boxy bulge' \citep{pohlen2004}.

%----------------------------FIG 2 
\begin{figure}[t]
\sidecaption[t]
%\texttt{B/P bulges seen in the edge-on view in the simulation models by Athanassoula et al. (2005, their figure 6): in the upper
%panel the line of sight is at 90$^{\circ}$ to the bar major axis, and in the 
%lower panel the B/P is seen end-on.}
% Use the relevant command for your figure-insertion program
% to insert the figure file.
% For example, with the option graphics use
\includegraphics[scale=.80]{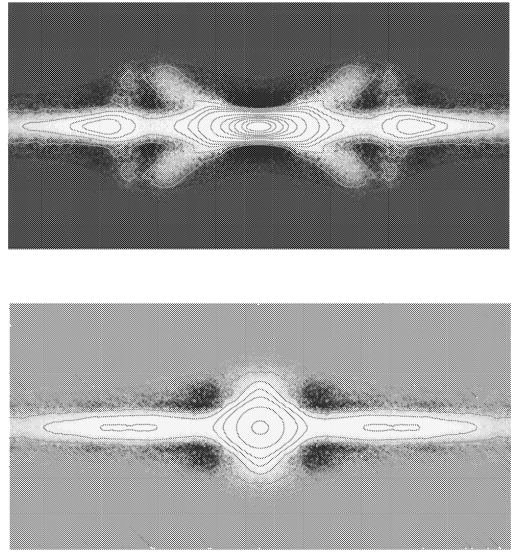}
%\includegraphics[scale=.65]{springer_fig3.pdf}
%
% If no graphics program available, insert a blank space i.e. use
%\picplace{5cm}{2cm} % Give the correct figure height and width in cm
%
\caption{Boxy/peanut bulges seen in the edge-on view in the simulation
  model by Athanassoula (2005, their figure 6): in the upper panel the
  line of sight is at 90$^{\circ}$ to the bar major axis, and in the
  lower panel the boxy/peanut is seen end-on.}
%\caption{If the width of the figure is less than 7.8 cm use the \texttt{sidecapion} comand \texttt{[t]} with the \texttt{sidecaption} command}
\label{fig:2}       % Give a unique label
\end{figure}

\section{Properties of the B/P bulges in the edge-on view}
\label{sec:3}

It was clear that more detailed analysis of the individual galaxies
were needed, either to prove or disprove the possible bar origin of
the B/P bulges. Such observations, particularly in the near-IR, where
the obscuring effects of dust are minimal, were carried out by several
groups. The observations were also compared with the predictions of
the simulation models.  

%-----------------------FIG 3 (NGC1055)
\begin{figure}[t]
%\sidecaption[t]
\includegraphics[scale=.65]{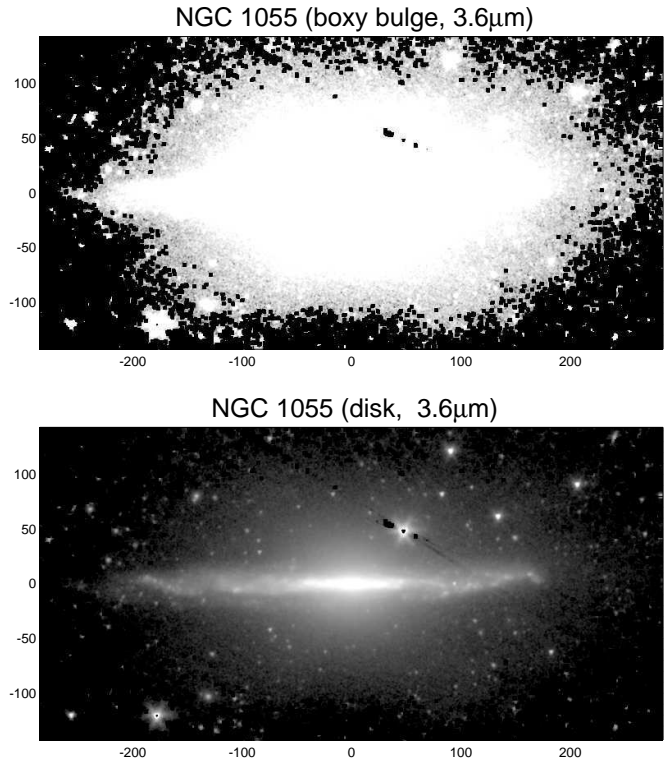}
\caption{NGC 1055, as an example of a thick boxy bulge. Shown 
is the 3.6$\mu$m Spitzer space telescope image in two 
different flux scales.  }
\label{fig:3}       % Give a unique label
\end{figure}
%-------------------------------------

\vspace{0.3cm}
\subsection{Direct images}

Morphology of the B/P-structures was systematically studied in the
near-IR by \citet{lutticke2000b}. They were able to show that a large
majority of these structures can be associated with bars.  In
particular, they emphasized that the B/Ps are not just thick bars, but
a combination of the vertically thick inner part of the bar, and a
central bulge formed differently. They showed that the degree of the
boxiness varies with the aspect angle of the bar, e.g. the bulge size
normalized with barlength correlates with the level of boxiness of the
bulge.  These results were consistent with the predictions of the
simulation models by \citet{pfenniger1991}, and they have been later
confirmed by other simulations
\citep{patsis2002,valpuesta2006,atha2002,debattista2005}.  L\"utticke,
Dettmar $\&$ Pohlen also measured the vertical thickness of the B/P,
both in respect of barlength, and the size of the bulge (e.g. the size
of the region with extra light above the exponential disk),
in agreement with the predictions of the simulation models.  The
measured size of the B/P, normalized to barlength was $\sim$0.4, again
in good agreement with the simulation models by Pfenniger $\&$
Friedli. Two possible explanations for the small central bulges were
speculated by \citet{lutticke2000b}: they could be
associated to the primordial bulges (i.e., 'classical bulges'), or to
the Inner Lindblad Resonance of the bar where a burst of star
formation increases the mass concentration (i.e., 'disky
pseudobulges'). This idea has been recently renovated by \citet{cole2014}.

\subsection{Unsharp masks}

The B/P bulges have been studied by \citet{aronica2003} using unsharp
mask images, which emphasize the sharp features that might appear in
galaxies.  They pointed out local surface brightness enhancements
along the bar major axis on both sides of the bar, which enhancements
appeared also in their simulation model after the bar had buckled.  To
illustrate this in Figure 4 (upper panel) a comparison is made between
a simulation model and the bar in ESO 443-042 observed at 3.6
$\mu$m. In the same figure we show also a typical barred early-type
galaxy, NGC 936, in almost face-on view.  Also this galaxy has flux
enhancements at the two ends of the bar, and it is tempting to argue
that they are the same features as in ESO 443-042.  We will come into
this issue later.

%--------------------------FIG 4b - X-shapes from Bureau2006
\begin{figure}[t]
%\sidecaption[t]
% Use the relevant command for your figure-insertion program
% to insert the figure file.
% For example, with the option graphics use
\includegraphics[scale=.35]{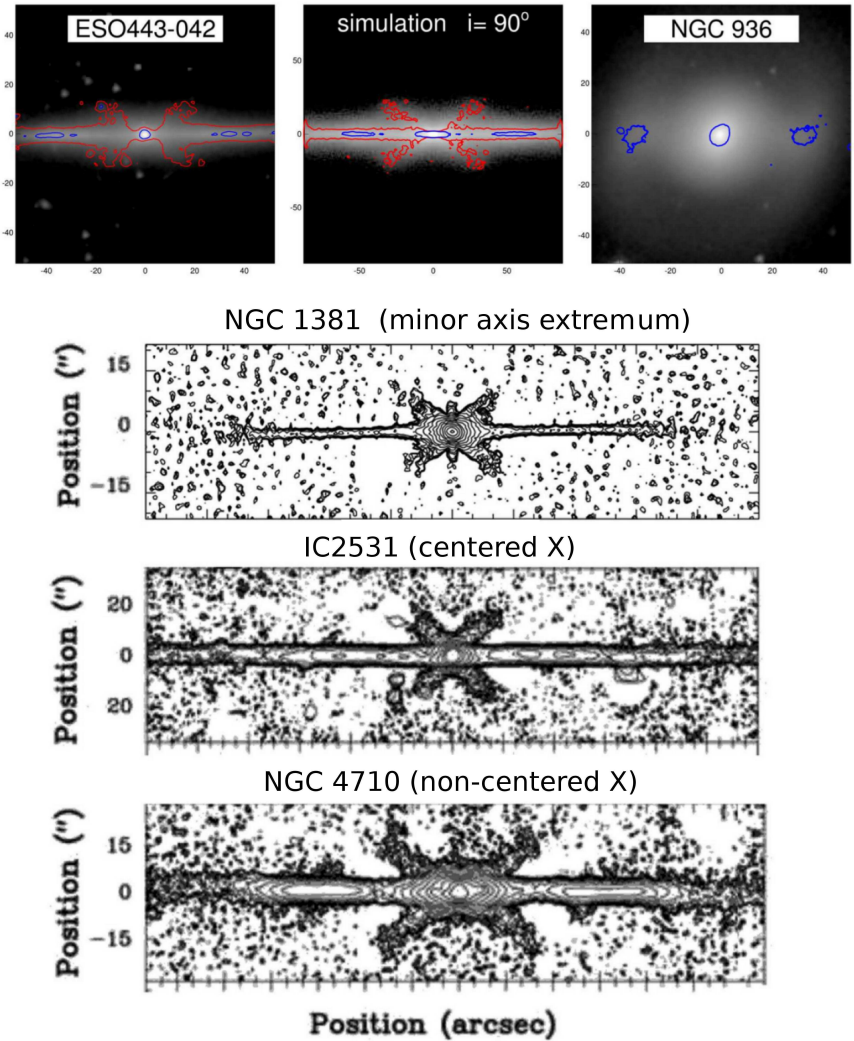}
%
% If no graphics program available, insert a blank space i.e. use
%\picplace{5cm}{2cm} % Give the correct figure height and width in cm
%
\caption{ The upper panel shows the Spitzer 3.6$\mu$m images for two
  galaxies with vertically thick inner bar components. For the edge-on
  galaxy ESO 443-042 it has an X-shape morphology, whereas in NGC 936
  it appears as a barlens. Also shown is a simulation model (the same
  as in Fig. 10) in which the bar has buckled in the vertical
  direction. The contours denote the unsharp masks of the same images:
  blue highlights the brightest regions and red color the
  X-shapes. The lower panels show different X-shape morphologies in
  the edge-on view, taken from \citet{bureau2006}.  }
%\caption{If the width of the figure is less than 7.8 cm use the \texttt{sidecap%ion} command to flush the caption on the left side of the page. If the figure is positioned at the top of the page, align the sidecaption with the top of the figure -- to achieve this you simply need to use the optional argument \texttt{[t]} with the \texttt{sidecaption} command}
\label{fig:4}       % Give a unique label
\end{figure}
%-----------------------

Unsharp masks were used to study the B/P bulges also by Bureau et
al. (2006), but now using a larger sample of 30 galaxies.  In fact, in the
study by Bureau et al. all the most important morphological
characteristics of the B/P/X-shape structures for the edge-on galaxies
are summarized, and are collected to our Figures 4 and 5:
 \vspace{0.3cm}

(a) {\it Secondary maximum appears along the bar major
axis} (upper row in Fig. 4), as discussed also by Aronica et al. (2003; see
also Patsis, Skokos $\&$ Athanassoula, their Fig. 5e).
\vspace{0.3cm}

(b) {\it The X-shapes can be centered or non-centered}, e.g. the
X-shape either crosses or does not cross the galaxy center (IC 2531
and NGC 4710 in Fig. 4).
\vspace{0.3cm}

(c) {\it Minor-axis extremum} appears (NGC 1381 in Fig. 4), e.g., there is a rather narrow and elongated local maximum in the
surface brightness profile along the minor axis near the galaxy center.
\vspace{0.3cm}

(d) {\it Spiral arms} start from the two ends of the B/P (lower left
panel in Fig. 5), e.g., symmetric, narrow, and elongated features
appear, which on the two sides of the B/P are shifted with each other.
\vspace{0.3cm}

%----------------------------FIG 5 - Erwin2013 vertailu
\begin{figure}[t]
%\sidecaption[t]
% Use the relevant command for your figure-insertion program
% to insert the figure file.
% For example, with the option graphics use
\includegraphics[scale=.70]{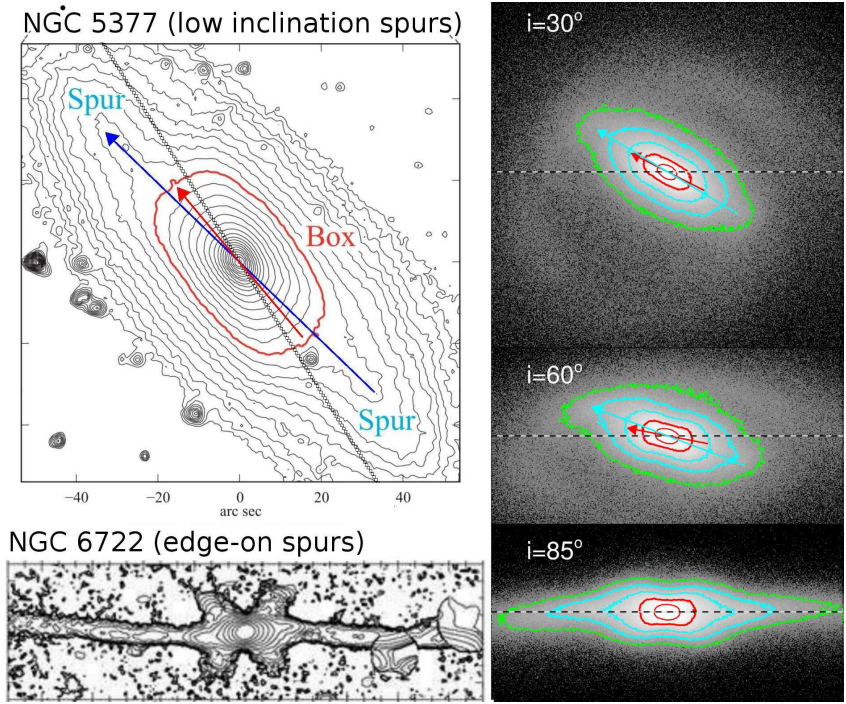}
%
% If no graphics program available, insert a blank space i.e. use
%\picplace{5cm}{2cm} % Give the correct figure height and width in cm
%
\caption{ 
Examples of barred galaxies with B/P-shape bulges at different viewing
  angles are shown. NGC 5377 shows isophotes from an archival Spitzer
  IRAC1 (3.6$\mu$m) image, and NGC 6722 is an unsharp mask of the
  K-band image. NGC 5377 has an inclination of 59$^{\circ}$, whereas
  NGC 6722 is almost edge-on.  The broad, nearly rectangular region in
  NGC 5377 is the boxy bulge, and the ``spurs'' projecting outside
  makes the outer part of the bar.  As an inclination effect, in both
  galaxies the ``spurs'' or ``spiral like features'', are twisted in
  respect to each other. Arrows mark their position angles, while the
  dotted line is the nodal line of the galaxy disk. Note how the
  position angle of the vertically thick boxy part falls between the
  position angles of the bar and the disk . Right panels show our
  simulations for a buckled bar (major axis makes and 35 degree angle
  with respect to nodal line), seen at different inclinations. The
  contours show the isophotes at different surface brightnesses,
  highlighting the inner and outer parts of the bar.
  }

\label{fig:5}       % Give a unique label
\end{figure}
%-------------------------------

The first two features are typical for the B/P bulges. According to
\citet{bureau2006} even 88$\%$ of the galaxies with B/Ps have a
secondary maximum along the bar major axis (in comparison to 33$\%$ in
their control sample). Also, 50$\%$ and 38$\%$ of the B/P bulges have
off-centered and centered X-shapes, respectively (in 33$\%$ in their
control sample with no B/P structures).  These features, as well as
the minor-axis extremum, have been predicted also by the simulation
models by \citet{atha2005}. Whether the X-shape is centered or not, in
the models depends on the azimuthal angle of the bar: when viewed
side-on the four branches of the X-feature do not cross the center.
However, for the centered X-shapes other explanations, like those
related to galaxy interactions, have also been suggested
\citep{binney1985,hernquist1988,whitmore1988}. Altogether, these
comparisons showed that the most important characteristics in the
morphology of the B/P bulges can be explained by bars.

Concerning the spiral arms, an overabundance in galaxies with B/P
bulges was found by \citet{bureau2006}. As a natural explanation for
that they suggested that the spiral arms are driven by bars, which is
indeed predicted also by the dynamical models. However, it has not
been convincingly shown how extended the bar driven spiral arms
actually are \citep[see][]{salo2010}. In fact, looking at the images
shown by Bureau et al. the structures that they call as spiral arms
are very similar to the features called as 'spurs' by
\citet{erwin2013} in their recent study of more face-on barred
galaxies (see upper left panel in Fig. 5).  The 'spurs' can be
understood as a combination of galaxy inclination, and the fact that
the inner and outer parts of the bar have different vertical
thicknesses. When the galaxy is not perfectly edge-on and the major
axis of the bar makes an angle with respect of the nodal line, 'spurs'
appear to be offset with respect to the major axis of the interior
isophotes associated with the boxy bulge (see Fig. 5, middle right
panel in our simulations). Using the words by Erwin $\&$ Debattista,
``in an inclined galaxy projection of the B/P creates boxy isophotes
which are tilted closer to the line of nodes than are the isophotes
due to the projection of the other, flat part of the bar, which form
the spurs''.  Taking into account that not all galaxies in the sample
by Bureau et al. are perfectly edge-on, in the two studies we are
obviously speaking about the same phenomenon.
 
\citet{lutticke2000b} left open the interpretation
of the central peaks in the surface brightness profiles of the B/P
bulges. However, \citet{bureau2006} take a stronger view stressing
that even the central surface brightnesses can be explained by the
processes related to the formation and evolution of bars. They argue
that classical bulges are not needed to explain their observations. As
a support for this interpretation Bureau et al. showed that the
surface brightness is more pronounced along the bar major axis than in
the azimuthally averaged brightness, which was suggested to mean that
most of the material at high vertical distances belongs to the B/P. In
their view the steep inner peak in the surface brightness profile
belongs to a flat concentrated inner disk (i.e., a 'disky pseudobulge'
in our notation). Alternatively, the central peak belongs to the bar,
formed as an inward push of the disk material when the bar was
formed. In principle the colors would distinguish between these
alternatives, but in the edge-on galaxies the central regions are
contaminated by dust and stellar populations of the outer disk.
 
%\vspace{0.3cm}
%FIG 5: Bureau+2006 (their fig 5)
\vspace{0.3cm}

\subsection{Structural decompositions}

Two edge-on galaxies with B/P bulges, NGC 4565 and NGC 5746, have been
decomposed into multiple structure components by \citet{kormendy2010}
and \citet{barentine2012}. In the
classification by \citet{buta2015} the Hubble types of these
galaxies are SB$_x$(r)ab sp, and (R')SB$_x$(r,nd)0/a sp.  In direct
infrared images the bulges in both galaxies clearly have boxy or even
X-shape morphology. The surface brightness profiles were decomposed
into an exponential disk, and two bulges (a 'boxy bulge' and a 'disky
pseudobulge') fitted with separate S\'ersic functions. In both galaxies
the boxy bulges were assumed to be bars seen in nearly end-on
view. For NGC 5746 there is also kinematic evidence for this
interpretation, manifested as a 'figure-of-eight' line-of-sight
velocity distribution, typical for boxy bars, which characteristic
will be discussed in more detail in the next section.

An interesting outcome of these decompositions is that most of the
bulge mass in these massive early-type disk galaxies resides in the
boxy bulge.  In NGC 4565 the boxy bulge-to-total mass ratio $B_{\rm
  boxy}/T$ $\sim$0.4, and $B_{\rm disky}/T$ $\sim$0.06 (i.e., a 'disky
pseudobulge' in our notation). The 3.6 and 8$\mu$m images and the
decomposition for this galaxy are shown in Figure 6.  Both type of
bulges are nearly exponential, along the major axis and perpendicular
to that. A more simple decomposition for the same galaxy by 
\citet{simien1986}, fitting only one de Vaucouleurs bulge
($n$=4) and an exponential disk, leads to $B/T$=0.4, which clearly
corresponds to that obtained for the 'boxy bulge' by Kormendy $\&$
Barentine. Although the relative mass of bulge is practically the
same in these two decompositions, the interpretation from the point
of view of galaxy formation is totally different.  This is one of
those cases where the simple decomposition approach finds a classical
bulge, although most of the bulge flux actually belongs to a boxy bar
component. A key issue in the structural decompositions is that
when there is a central peak in the surface brightness profile, and
the various components of the disk are not fitted separately, 
leads to a massive bulge with large S\'ersic index, typical for
classical bulges.

Although a systematic study of the decompositions for B/P bulges seen
in the edge-on view is still needed, the above discussed
decompositions have already shown that there exist massive early-type
spiral galaxies which have no classical bulges. Or, at least it is not
self-evident how these tiny exponential central bulges should be
interpreted.

%------------------------FIG 6 - Kormendy decomposition
%Figure 5

\begin{figure}[t]
%\sidecaption[t]
% Use the relevant command for your figure-insertion program
% to insert the figure file.
% For example, with the option graphics use
\includegraphics[scale=.80]{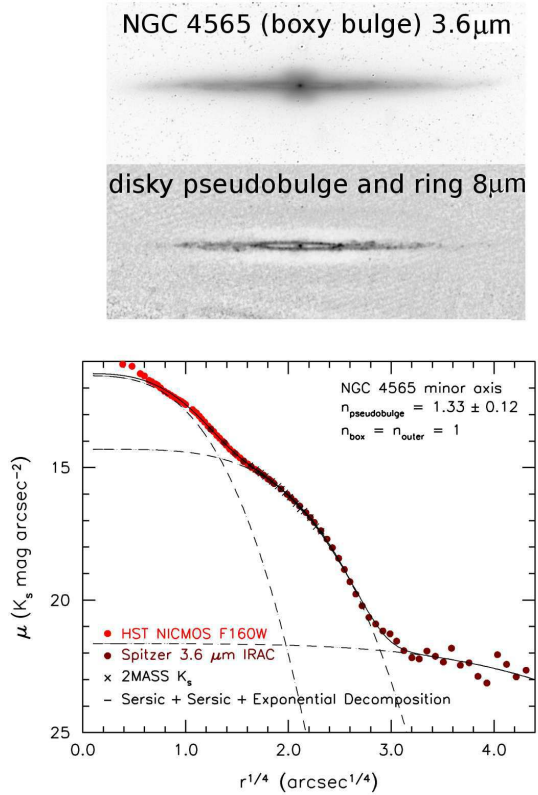}
%
% If no graphics program available, insert a blank space i.e. use
%\picplace{5cm}{2cm} % Give the correct figure height and width in cm
%
\caption{ Spitzer/IRAC 3.6$\mu$m and 8$\mu$m images of NGC 4565 (upper
  and middle panels) shown to emphasize the boxy bulge, the ring and
  the tiny central pseudobulge.  The lower panel shows a composite
  minor axis surface brightness profile made of the 3.6$\mu$m image
  (brown points), and the Hubble space telescope image at F160W band
  (red points). The central pseudobulge and the boxy bulge are fitted
  with S\'ersic functions and the outer structure with an exponential
  function. The solid line is the sum of the components.  The nucleus
  is not fitted.  With permission the figure is taken from 
 \citet{kormendy2010}.  }

\label{fig:6}       % Give a unique label
\end{figure}

%\clearpage
%---------------------------Fig. 7
%\vspace{0.3cm}

\begin{figure}[t]
%\sidecaption[t]
% Use the relevant command for your figure-insertion program
% to insert the figure file.
% For example, with the option graphics use
\includegraphics[scale=.80]{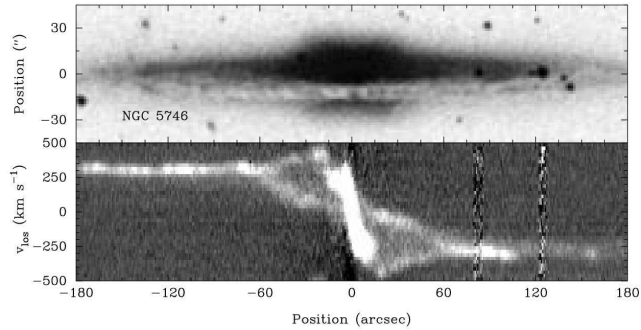}
%
% If no graphics program available, insert a blank space i.e. use
%\picplace{5cm}{2cm} % Give the correct figure height and width in cm
%
\caption{ The upper panel shows the K-band image of NGC 5746, and the
  lower panel the ionized gas [NII]6584 emission line position
  velocity diagram taken along the major axis. 
  With permission the
  figure is taken from \citet{bureau1999}.  }

\label{fig:7}       % Give a unique label
\end{figure}

%\clearpage
%--------------------------------------------Fig. 8
\begin{figure}[t]
%\sidecaption[t]
% Use the relevant command for your figure-insertion program
% to insert the figure file.
% For example, with the option graphics use
\includegraphics[scale=.80]{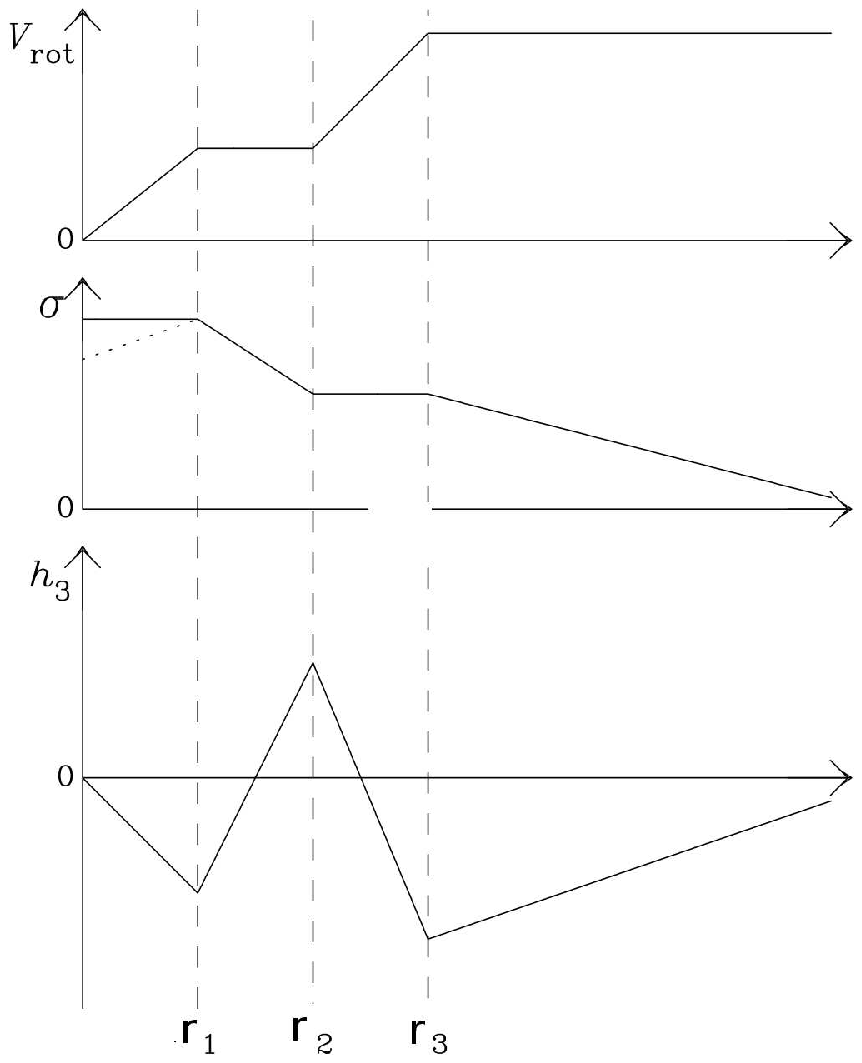}
%
% If no graphics program available, insert a blank space i.e. use
%\picplace{5cm}{2cm} % Give the correct figure height and width in cm
%
\caption{ 
  Schematic view of the main diagnostic tools to identify bars and
  B/P bulges.  Shown are the radial profiles of the rotation velocity
  V$_{rot}$, the stellar velocity dispersion $\sigma$, and third
  Gauss-Hermite moment parameter h$_3$. The vertical dashed lines
  indicate the radii of the central disk pseudobulge ($r_1$), boxy
  bulge ($r_2$) and the whole bar ($r_3$). With permission the
  figure is taken from \citet{bureau2006}.  }

\label{fig:8}       % Give a unique label
\end{figure}

\vspace{0.3cm}
\subsection{Diagnostics of bars in gas and stellar kinematics}  
\vspace{0.3cm}

Although the B/P morphology can be easily recognized in the edge-on
view, in most cases the images alone cannot tell whether those
structures indeed form part of the bar or not. Bars in the edge-on
view can be easily mixed with rings or lenses. Since the components
might have different velocity dispersions which can be hotter than the
underlying disk, kinematic tools to identify bars are important.

One such tool suggested by Kuijken $\&$ Merrifield (1995, see also Vega
Beltran et al. 1997), is the 'figure-of-eight' structure in the
line-of-sight velocity distribution (LOSVD) of galaxies, derived from
the emission or absorption lines.  It is based on the idea that the
LOSVD has two peaks, one due to particles traveling in bar-related
stellar orbits faster than the local circular velocity, and particles
that travel more slowly than that. Variations in these velocities then
form the 'figure-of-eight' in the diagram where the velocity is shown
as a function of galaxy radius (see Fig. 7).  A good correspondence of
these observations with the predictions of the simulation models was
obtained by \citet{atha1999}, but soon it also became clear that this
method works only for strong bars \citep{bureau2005}.

The diagnostic tools were further developed by Chung $\&$ Bureau
(2004, see also the review by Athanassoula 2005), with the main
emphasis to distinguish, not only bars in general, but also the B/P
bulges within the bars. The identification is based on inspection of
the velocity ($V_{rot}$), the stellar velocity dispersion ($\sigma$),
and the third and forth terms of the Gauss-Hermite parameters along
the major axis \citep[see][]{{bender1994}}, which measure the asymmetric
($h_3$) and symmetric ($h_4$) deviations of the LOSVD from a pure
Gaussian. The $h_3$ parameter is expected to be a good tracer of the
triaxiality of the bulge. The main diagnostics of bars with B/P bulges
are shown in Figure 8. They show 'double humped' rotation curves,
flat-top or weakly peaked $\sigma$-profile, and that $h_3$-profile correlates
with $V_{rot}$ over the projected bar length. Also $h_4$-profile, although
being a weaker indice, shows
central and secondary minima.  \citet{chung2004} studied 30
edge-on spirals (24 with B/Ps) and showed that even 90$\%$ of them
showed kinematic signatures of bars. Not only bars were identified,
but also the edges of the B/Ps were recognized in the $h_3$-profiles.
A large fraction (40$\%$) of those galaxies also showed a drop in the
central velocity dispersion, being a manifestation of a dynamically
cold central component.

\vspace{0.3cm}

%\subsection{Kinematic Diagnostics tools}

\section{Detection and properties of B/P-shape bulges in face-on systems}
\label{sec:4}

In face-on view the problem is the opposite: bars are easy to
recognize in the images, but the B/P/X-shape structures, which are
assumed to be thick in the vertical direction, presumably disappear in
the face-on view. In fact, excluding the edge-on galaxies, the
B/P-shape structures were expected to be visible only in a narrow range
of galaxy inclinations near to the edge-on view. Nevertheless, using
the words by \citet{kormendy2010}: ``as long as face-on and
edge-on galaxies appear to show physical differences we cannot be sure
that we understand them.''

\subsection{Isophotal analysis}

%What do we know about the B/P/X-shape bulges in the more face-on view?
Isophotal analysis of the image contours has actually appeared to be a
powerful tool to identify B/Ps in moderately inclined galaxies.  This
has been shown in a clear manner by \citet{beaton2007} for M31 (see
also Athanassoula $\&$ Beaton 2006 for simulations), which galaxy has
an inclination of 77.5$^{\circ}$. The main idea is to fit ellipses to
isophotes, and to measure the deviations from the elliptical
shapes. The sine ($A_4$) and cosine ($B_4$) terms of the Fourier
series measure the boxiness and diskiness of the isophotes (see
Fig. 9). In the boxy region $B_4$ is positive and $A_4$ is negative.
Characteristic for the boxy region is also that the ellipticity
increases towards the edge. On the other hand, the position angle is
maintained constant throughout the bar region, at least for strong
bars. The image of M31 is not shown here, but it would look very much
like NGC 5377 in our Figure 5, which galaxy has boxy inner isophotes
associated to the boxy bulge, and 'spurs' associated to the more
elongated part of the bar.
 
A similar analysis for a larger number of galaxies has been made by
\citet{erwin2013}. They studied 78 barred S0-Sb galaxies, covering a
large range of galaxy inclinations.  The leading idea in their study
was to find out an optimal range of galaxy inclinations
(i$<$45$^{\circ}$) and the bar's position angles from the nodal line
for the detection of B/P.  Using a small parameter space they were
able to study galaxies in which both the B/P bulge and the large scale
bar could be identified in the same galaxies. This allowed also a more
reliable estimate for the relative size of the B/P-structure ($R_{\rm
  boxy}$/$R_{\rm bar}$ $\sim$ 0.4), which appeared to be similar to
that predicted by the simulation models of bars
\citep{pfenniger1991,atha2002,debattista2005}, and is also similar to
those obtained by \citet{lutticke2000b} in observations. Also,
extrapolating the number statistics of B/P bulges found in the ideal
range of all bar/disk orientations and galaxy inclinations, they
estimated that even 2/3 of bars might have B/P bulges. Taking into
account that a certain fraction of bars at all inclinations must be
end-on, this fraction is not far away from the suggestion made by
\citet{lutticke2000b} that all bars might have B/P
structures. However, the extrapolation made by Erwin $\&$ Debattista
for making their prediction is based only on a few galaxies with
identified bars and B/P structures in the same galaxies.

A large majority of bars in the sample by \citet{erwin2013}
have boxy, rather than peanut-shape isophotes. A given explanation was
that in the central regions of the B/P-structures there exist extra
inner disks or compact bulges, which smooth out the peanut shape.  In
fact, the basic assumption in all the morphological and isophotal
analysis of the B/P bulges discussed above is that the vertically
thick inner parts of bars have either boxy or peanut shapes.  However,
based on the simulation models by Athanassoula et al. (2014) 
that is not necessarily the case in the face-on view where they 
can appear fairly round. In fact, the orbits populating the bars
might have more complicated structures than just regular orbits
around 3D bar-supporting periodic orbits \citep[see][]{patsis2014a}.

%---------------------------------FIG 9 - Beaton+
\begin{figure}[t]
%\sidecaption[t]
% Use the relevant command for your figure-insertion program
% to insert the figure file.
% For example, with the option graphics use
\includegraphics[scale=.85]{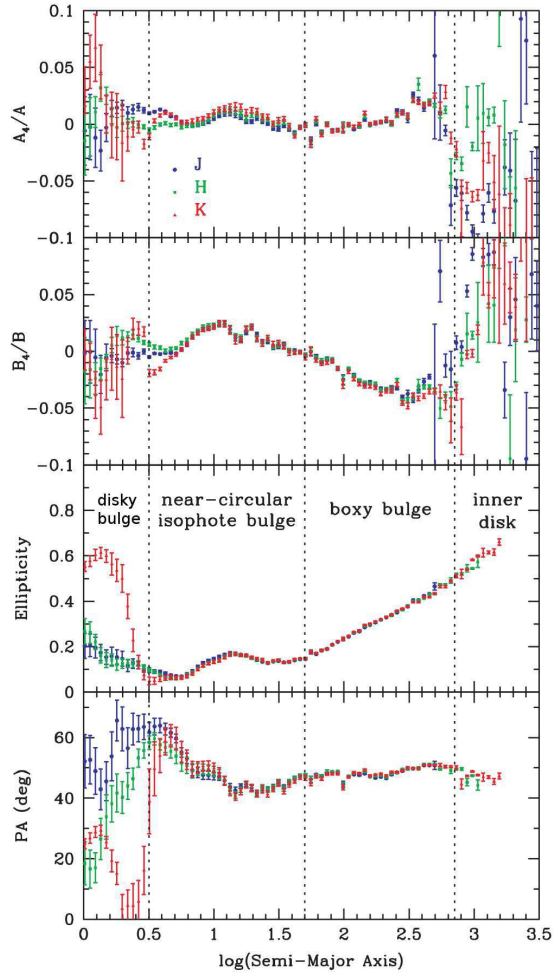}
%
% If no graphics program available, insert a blank space i.e. use
%\picplace{5cm}{2cm} % Give the correct figure height and width in cm
%
%\caption{Please write your figure caption here}
\caption{
Isophotal analysis of M31 which galaxy has a boxy bulge. Radial
profiles (in arcseconds) of the ellipticities and position angles of
the isophotes are shown in the two lower panels. The upper panels show
the sine ($A_4$) and cosine ($B_4$) terms of the Fourier series of the same
isophotes.  Shown separately are the measurements in J, H and K-band
bands. The regions covering the near-nuclear bulge ('disky pseudobulge'
in our notation), boxy bulge, and inner disk (equivalent to 'spurs' in
our Figure 5) are shown by dotted lines. With permission the figure is 
taken from \citet{beaton2007}. 
}
\label{fig:9}       % Give a unique label
\end{figure}
%---------------------------------------

\subsection{Properties of barlenses}

A different approach was taken by \citet{lauri2011} who identified
distinct morphological structures called barlenses, in a sample of
$\sim$200 early-type disk galaxies (NIRS0S atlas), observed at fairly
low galaxy inclinations ($i\le 65^\circ$).  Barlenses were recognized
as lens-like structures embedded in bars, covering typically half of
the barlength. In distinction to nuclear lenses they are much larger,
and compared to classical bulges the surface brightness distribution
decreases much faster at the edge of the structure. In
\citet{lauri2011} and in \citet{buta2015} barlenses have been coded
into the classification. These structures (though not yet called as
such) were decomposed with a flat Ferrers function in many S0s already
by \citet{lauri2005}. This kind of decompositions were summarized in
\citet{lauri2010}. In \citet{lauri2007} it was speculated that such
inner lens-like structures might actually be the face-on views of the
vertically thick B/P bulges. If barlenses indeed are physically the
same phenomenon as the B/P bulges, that would make possible to have a
consistent view of the relative masses of the classical and
pseudobulges at all galaxy inclinations.

Two prototypical barlens galaxies, NGC 936 and NGC 4314, are shown in
Figures 4 and 10. In the images barlenses can be easily mixed with the
classical bulges. However, in the surface brightness profiles
barlenses appear as nearly exponential, flat sub-sections, both along
the major and the minor axis of the bar. Characteristic morphological
features for barlens galaxies are the ansae (or handles), which appear
at the two ends of the bar (see NGC 936 in Fig. 4). It has been shown
\citep{lauri2013} that even half of the barlenses are embedded in that
kind of bars. In Section 3.2 we discussed that such flux enhancements
are produced also in galaxy simulations, at the same time when the bar
buckles in the vertical direction. This can be considered as further
indirect evidence supporting the idea that barlenses indeed form part
of a buckled bar.  The unsharp mask image of NGC 4314 also shows a
structure connecting the barlens to the more elongated part of the bar
(see Fig. 10).  Using the measurements in the NIRS0S atlas
\citet{atha2014} showed that, in respect of the bar, barlenses have
very similar sizes as obtained for the B/P bulges by
\citet{lutticke2000b} and \citet{erwin2013}.

Morphological structures similar to the observed barlenses are
produced by N-body and smoothed particle hydrodynamical simulations by
\citet{atha2013}. In \citet{atha2014} more comparisons between the
observations and models are shown. In their models barlenses appear in
the face-on view without invoking any spheroidal bulge components in
the initial models. An example of a barlens in such simulation model,
seen both in the edge-on and face-on views, is shown in Figure 10 (two
lower left panels).  Recent orbital analysis of bars by
\citet{patsis2014b} have shown that ``sticky chaotic'' orbits,
building parts of bars can appear at high vertical distances in such a
manner that when seen in the face-on view they form a boxy inner
structure inside the bar (their fig. 8). These bar orbits might be the
ones associated to barlenses in some cases. Whether also the X-shape
is visible inside the boxy component depends on the specific
combination of the orbital families of bars.

\subsection{Barlenses - the face-on counterparts of B/P bulges}

If barlenses and B/P/X-shape bulges indeed were physically the same
phenomenon, just seen at different viewing angles, we should see that
in the number statistics in a representative sample of nearby
galaxies. That has been looked at by \citet{lauri2014} using a sample
of 2465 nearby galaxies at 3.6$\mu$m or 2.2$\mu$m wavelengths,
covering all Hubble types and galaxy inclinations (a combination of
NIRS0S, and the Spitzer Survey of Stellar Structure of galaxies
S$^4$G). In order to find out all the X-shape structures unsharp masks
were done for all these galaxies, in a similar manner as was done
previously by \citet{bureau2006} for a representative sample of
edge-on galaxies.  Barlenses were recognized in the galaxy
classifications of \citet{buta2015} and \citet{lauri2011}.
Remarkably, the apparent axial ratios of the galaxies with barlenses
and X-shape structures are consistent with a single population viewed
from random orientations (Fig. 11, upper panel). Although barlenses
appear in less inclined galaxies, there is a large overlap in their
parent galaxy inclinations, compared to those with X-shape
structures.  The parent galaxies of barlenses and X-shape structures
have similar distributions of total stellar mass (Fig. 11, lower
panel), and also similar red colors.  It is worth noticing that the
peak in the mass distribution of these galaxies appears at the Milky
Way mass.  There are also similarities in the kinematics of barlenses
and X-shape structures, which will be discussed in Section 4.5.

It appears that among the S0s and early-type spirals even half of the
barred galaxies have either a barlens or an X-shape structure, and
$\sim$30$\%$ if also the non-barred galaxies are included in the
statistics \citep{lauri2014}. This is not much less than the
45$\%$ of B/Ps found by \citet{lutticke2000a} among
the edge-on galaxies in the same morphological type bin.  The slightly
lower B/P fraction by Laurikainen et al. can be explained by the fact
that limiting to X-shape structures, most probably they picked up only
the strong bars where the X-shapes are more pronounced
\citep{atha2005}. As L\"utticke, Dettmar $\&$ Pohlen did not use
any unsharp masks we don't know how many of the galaxies in their
sample actually have X-shape structures. Most probably not all of
them, because boxy isophotes can be identified in the edge-on view
even if the bulges have no X-shapes.  Fractions of B/Ps has been
recently studied also by \citet{yoshino2014} in a sample of
1700 edge-on galaxies in the optical region. In order to identify bars
a comparison sample of 2600 more face-on galaxies was used. It was
then assumed that the bar fraction is the same among the edge-on
galaxies. They found that B/Ps appear in 20$\%$ of the galaxies, which
fraction is much lower than the 45$\%$ found by \citet{lutticke2000a}.
However, according to Yoshino $\&$ Yamauchi the
fraction of B/Ps they found is very similar to that obtained by
L\"utticke, Dettmar $\&$ Pohlen if the weakest category of B/Ps by
L\"utticke et al. is omitted.

%------------------------------FIG 10 - barlens obs/model
%\vspace{0.3cm}
%Fig. 9: obs vs. model of barlens gal (L+2014, fig. 1) 
%\vspace{0.3cm}

\begin{figure}[t]
%\sidecaption[t]
% Use the relevant command for your figure-insertion program
% to insert the figure file.
% For example, with the option graphics use
\includegraphics[scale=.55]{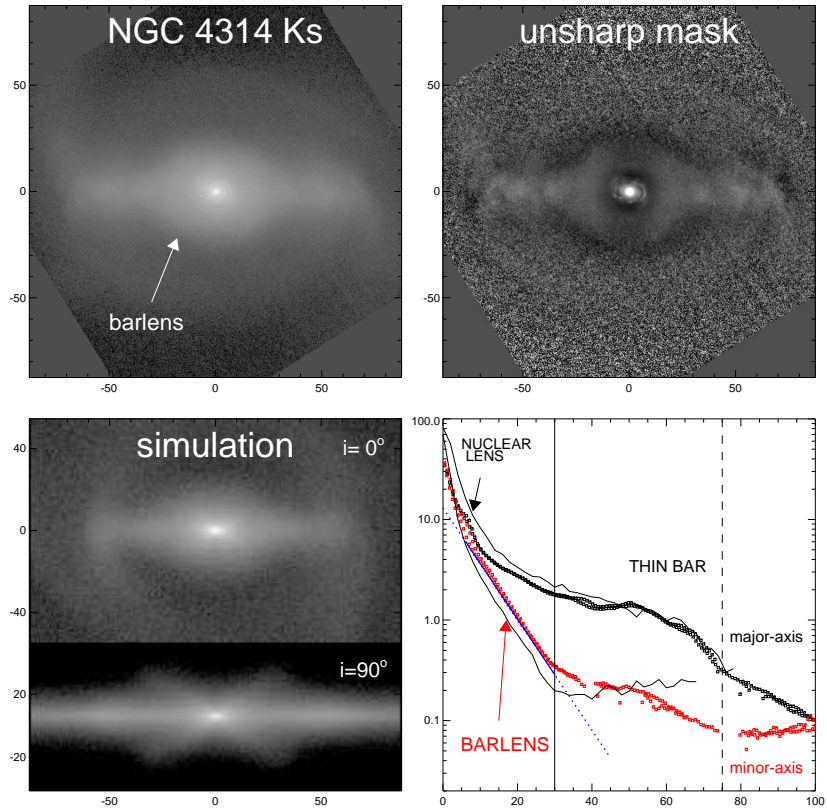}
%
% If no graphics program available, insert a blank space i.e. use
%\picplace{5cm}{2cm} % Give the correct figure height and width in cm
%
%\caption{Please write your figure caption here}
\caption{
An example of a barlens galaxy NGC 4314, showing the K$_s$-band image
   from Laurikainen et al. (2011; upper left panel) and the unsharp
   mask image of that (upper right panel). The surface brightness
   profiles along the bar major (black symbols) and minor axis (red
   symbols) is also shown.  The lower left panel shows the simulation
   model gtr115 from \citet{atha2013,atha2014}, both in the
   face-on and edge-on view. The simulation model profiles are shown
   by solid lines in the profile plot.  Axis labels are in arcseconds
   in all panels.  With permission the figure is taken from
   \citet{lauri2014}.
}
\label{fig:10}       % Give a unique label
\end{figure}

%----------------------------------FIG 11 - inc-distr
%\vspace{0.3cm}
%Fig 10: INC-distribution of bl+X, L+2014
%\vspace{0.3cm}

\begin{figure}[t]
%\sidecaption[t]
% Use the relevant command for your figure-insertion program
% to insert the figure file.
% For example, with the option graphics use
\includegraphics[scale=.70]{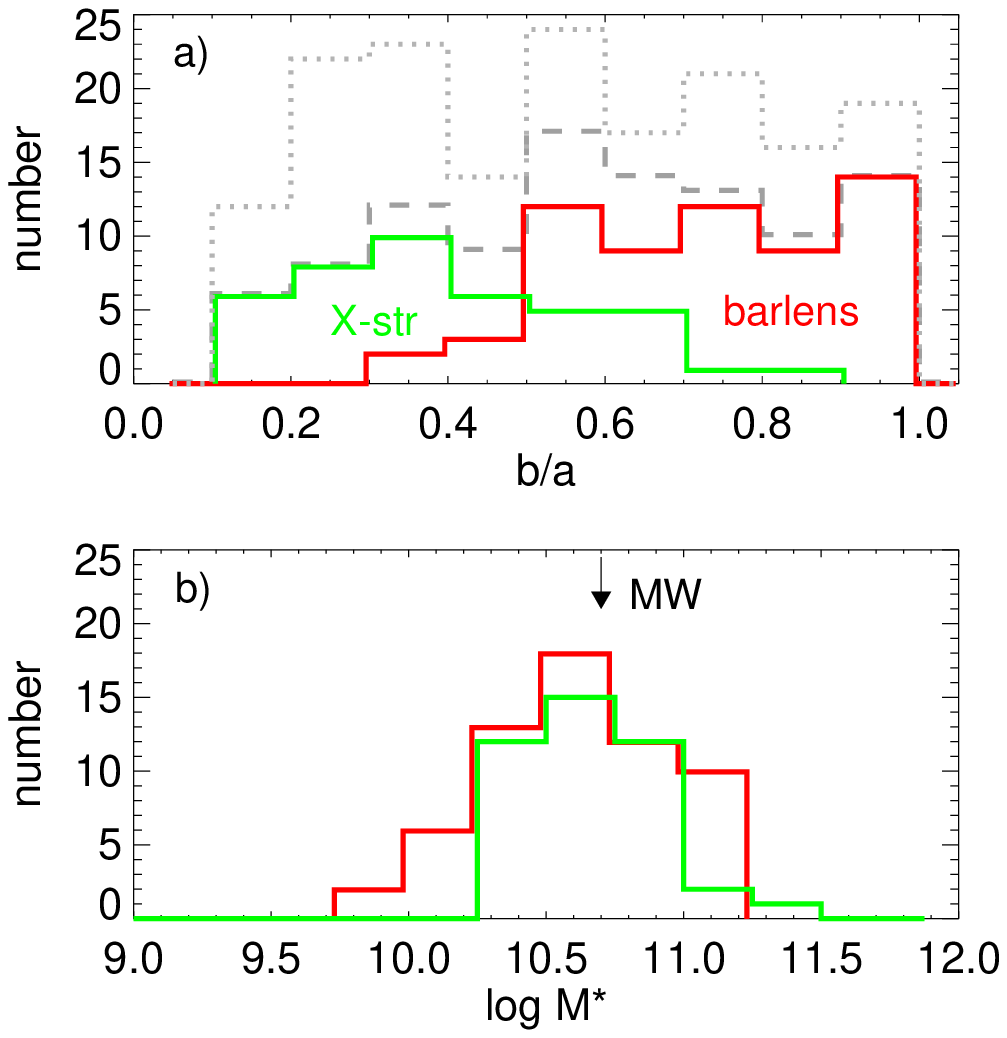}
%
% If no graphics program available, insert a blank space i.e. use
%\picplace{5cm}{2cm} % Give the correct figure height and width in cm
%
\caption{
The distributions of (a) galaxy minor-to-major ($b/a$) axis ratios and (b)
   total stellar masses (in units of solar masses) of  the
   galaxies, hosting either barlens (red) or X-shape structures
   (green). In these plots a magnitude-limited ($B_T \leq$12.5 mag)
   sub-sample of 365 barred galaxies in a combined S$^{4}$G and NIRS0S is used.
The grey dashed and dotted lines show the barlens and X-shapes structures,
in the magnitude-limited sample and the complete S$^{4}$G, respectively. 
}

\label{fig:11}       % Give a unique label
\end{figure}
%--------------------------------------

\subsection{Structural decompositions}

The assumption that barlenses and B/P bulges are physically the same
phenomenon allows us to estimate the relative masses of these
components, since this can be done in a fairly reliable manner at
moderate galaxy inclinations ($i\le 65^\circ$) using multi-component
decompositions. When the inclination of the disk increases, the
reliability of these mass estimates rapidly decreases. The
decompositions of \citet{barentine2012} discussed earlier were made
to one-dimensional surface brightness profiles, which is indeed a
reasonable approach for the galaxies in the edge-on view. However,
applying a similar approach in a more face-on view, in particular when
the bar has two components, would dilute the non-axisymmetric
structure components, which would appear as one big bulge in the
average surface brightness profile (in terms of the flux above the
disk). A better approach is to fit the two-dimensional flux
distributions of the galaxies.
 
Examples of the decompositions using a two-dimensional approach and
fitting the two bar components separately, allowing also the
parameters of the B/P/barlens to vary, are taken from \citet{lauri2014}.
They used the 3.6$\mu$m Spitzer images to decompose 29
nearby galaxies having either a barlens or an X-shape structure. The
bulges (i.e., the central mass concentrations) and disks were fitted
with a S\'ersic function, whereas the two bar components were fitted
either using a Ferrers or a S\'ersic function. Representative examples
of these decompositions are shown in Figure 12.  It appeared that the
relative fluxes of barlenses and X-shape bulges form on average even
10-20$\%$ of the total galaxy flux, in comparison to $\sim$10$\%$ in
the central bulges (i.e., a 'disky pseudobulges'). In IC 5240 (Sa) the only
bulge seems to be the X-shaped bar component. On the other hand, NGC
4643 (S0$^+$) might have also a small central bulge embedded in the
barlens. Looking at the surface brightness profile alone we don't know
for sure whether the central peak is really a distinct bulge component, or is it
rather formed of the same material as the rest of the bar, at the
epoch of bar formation. The S\'ersic index of the central component is
$n$=0.7 indicating that it is not a classical bulge. One possibility
is that it is a manifestation of an old pseudobulge formed at high
redshift, composed of old stars. That kind of pseudobulges form in
the hydrodynamical cosmological simulations by \citet{guedes2013}, 
via a combination of disk instabilities and minor mergers.

If the B/P bulges (i.e., the barlenses or X-shape structures in the
above decompositions) are omitted in the decompositions that would
dramatically affect the obtained relative masses of the classical
bulges. In the early-type disk galaxies the deduced central bulge will
increase from 10$\%$ to 35$\%$ in the sample by \citet{lauri2014}, the
value 35$\%$ being consistent with the previous more simple
bulge/disk/bar decompositions (Gadotti 2009, Weinzirl et al. 2009; see
also the more simple decompositions by Laurikainen et al. 2006).
In section 3.3 we
discussed an edge-on galaxy, NGC 4565, for which galaxy the same
happens when the simple and more detailed decompositions are compared.

We can compare the decompositions by \citet{lauri2014} with
those obtained by \citet{erwin2003} for NGC 2787 and NGC 3945. Using
a completely different decomposition approach they ended up with 
similar small relative masses for the central bulges in these two
galaxies, just different names were used for the bulges. What is a
barlens in \citet{lauri2014}, is called as an ``inner disk''
by Erwin et al., which disks are a magnitude larger than the central
classical bulges, manifested as peaks in the surface brightness
profiles.  Similar approach as in \citet{erwin2003} has been
recently taken also by \citet{erwin2015} for additional seven barred
early-type galaxies, but now calling the ``inner disks'' as ``disky
pseudobulges''.  They identified boxy isophotes only in one of those
galaxies, but many of them are classified as having a barlens by
\citet{lauri2011}.

The small central bulges in barred galaxies with B/P bulges are indeed
intriguing, and the obtained nature of those bulges depends on how they
are interpreted. As discussed by \citet{chung2004}, and more
recently by \citet{abreu2014}, cool disk components
manifesting as nuclear rings or spiral arms, are often embedded in the
B/P bulges. An example of a barlens galaxy with similar
characteristics is NGC 4314, showing a star forming nuclear ring
inside the boxy bulge. However, at least in the near-IR these central
cool components are expected to contribute very little in the surface
brightness profiles.

%------------------FIG 12 - decomp
%\vspace{0.3cm}
%Fig 11:  deco,mpositions from L+2014
%\vspace{0.3cm}

\begin{figure}[t]
%\sidecaption[t]
% Use the relevant command for your figure-insertion program
% to insert the figure file.
% For example, with the option graphics use
\includegraphics[scale=.50]{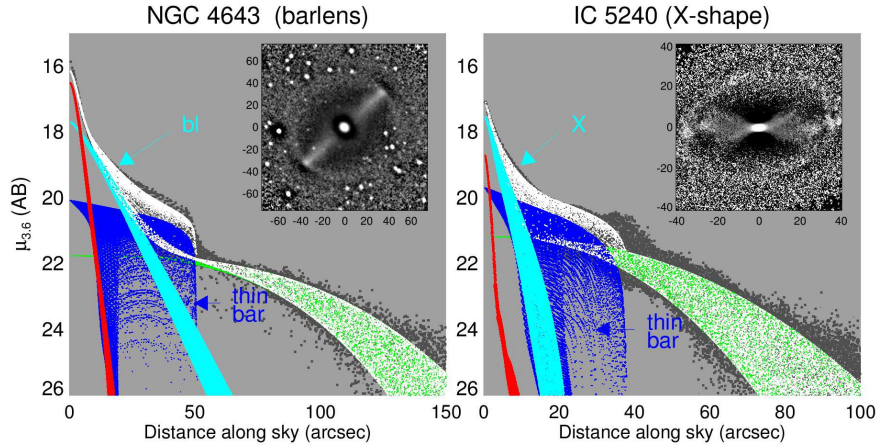}
%
% If no graphics program available, insert a blank space i.e. use
%\picplace{5cm}{2cm} % Give the correct figure height and width in cm
%
\caption{ Decomposition models for the barlens and X-shaped galaxies
  NGC 4643 and IC 5240, which are barred galaxies with Hubble stages
  S0$^+$ and Sa, respectively. The unsharp mask images are shown in
  the small inserts in the upper corners.  Black dots are the pixel
  values of the two-dimensional flux-calibrated 3.6$\mu$m Spitzer
  images, and white dots show the pixel values of the total
  decomposition models. Red and green dots show the bulge and the disk
  components, whereas the dark and light blue indicate the thin and
  thick bar (i.e., the barlens and X-shape structure) components. With
  permission the figure is taken from \citet{lauri2014}.  }

\label{fig:12}       % Give a unique label
\end{figure}

\subsection{Diagnostics of B/P bulges of stellar kinematics}

Using simulation models an attempt to identify B/P structures in more
face-on galaxies was done by \citet{debattista2005}. The diagnostics
largely relies on the analysis of the forth-order Gauss-Hermite
moment, $h_4$ along the bar major axis. A B/P bulge is recognized as
negative double minima in the $h_4$-profile. These minima appear
because the vertical velocity distribution of stars becomes broader,
for which reason $h_4$ is a good proxy for the unobservable vertical
density distribution.  However, in spite of the smart idea, this method
has been applied only for a few galaxies with B/P bulges, like NGC 98
\citep{abreu2008}. The reason is that the observations are
very demanding and require a large amount of observing time at large
telescopes.  The diagnostics for NGC 98 is shown in Figure 13. The
size of the B/P is estimated from the radius of the minimum in the
$h_4$-profile, which in this case is 0.35 times the bar semi-major axis
length. The same diagnostics has been recently applied for ten more
face-on barred galaxies by \citet{abreu2014}. They
identified B/P bulges in two additional galaxies, and marginally in
three more galaxies.  In four of these galaxies a dynamically cool
central component inside the B/P bulge was the only central bulge,
without any sign of a dynamically hot classical bulge.

\vspace{0.3cm}

%-------------------------------------Fig. 13
\begin{figure}[t]
\sidecaption[t]
% Use the relevant command for your figure-insertion program
% to insert the figure file.
% For example, with the option graphics use
\includegraphics[scale=.80]{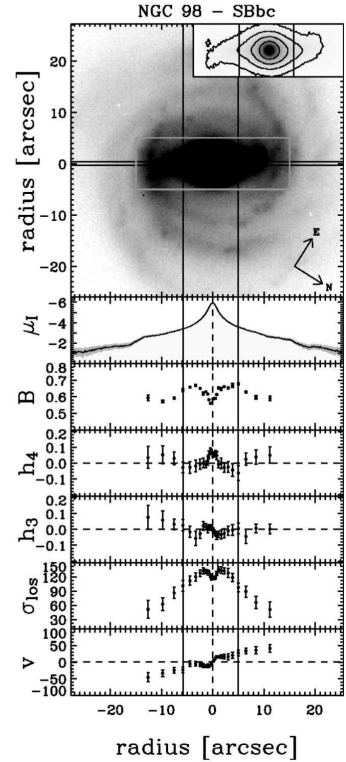}
\caption{ Morphology and stellar kinematics of NGC 98.  The top panel
  shows the I-band image, where the slit position and image
  orientation are also indicated. The inset shows the portion of the
  galaxy image marked with a white box. The the panels show from top
  to bottom the radial profiles of surface brightness, broadening
  parameter B, forth ($h_4$) and third ($h_3$) moments of the
  Gauss-Hermite series, line-of-sight velocity dispersion
  ($\sigma_{los}$), and the stellar velocity v, The two vertical lines
  indicate the location of the $h_4$ minima associated with the B/P
  region in NGC 98. The figure is from \citet{abreu2008}.
}
\label{fig:13} 
\end{figure}
\vspace{0.3cm}

In the above studies long-slit spectroscopy was used.  In principle
integral-field unit (IFU) spectroscopy would be ideal to trace the
B/P/X/barlens features, but the resolution and the field-of-view have
not yet been sufficient for very detailed studies of B/Ps (see the
review by Falcon-Barroso in Section 3). So far the largest survey
using IFU spectroscopy is the Atlas$^{3D}$, in which 260 nearby early-type
galaxies, at all galaxy inclinations, have been mapped within one
effective galaxy radius \citep{cappellari2007,emsellem2011}.  Almost
all bulges (86$\%$) were found to be fast rotating, which is
consistent with the idea that most of the bulge mass in the nearby
galaxies resides in the B/P/barlens bulges, rotating with the
underlying disk \citep[see][]{lauri2014}.  Most probably due to the
limitations of the observations only a small fraction (15$\%$) of the
fast rotating bulges showed signatures of B/P-structures, in terms of
double peaked rotation curves or twisting isophotes
\citep[see][]{krajnovic2011}.

We used the Atlas$^{3D}$ in the following manner.  We picked up all those
barred galaxies that have either a barlens or an X-shaped structure
identified in the images in the sample by \citet{lauri2014},
and then looked at what kind of kinematics the Atlas$^{3D}$ finds for those
galaxies.  We found 27 galaxies in common between the two surveys (11
with X-shapes, 16 with barlenses). It appeared that all these galaxies
are classified as regular fast rotators in \citet{emsellem2011}.
Half of them have 'double humped' rotation curves,
indicative of B/P-structures, but the other half has no particular
kinematic features of which the vertically thick bar components could
be identified.  It is interesting that the fractions of the double
humped rotation curves are fairly similar among the galaxies having
barlenses and X-shaped structures (56$\%$ and 36$\%$, respectively),
which fractions are much higher than for the bulges in the Atlas3D in
general.  This is consistent with the idea that barlenses and B/P
bulges are manifestations of the same physical phenomenon.

From the point of view of galaxy formation, the interpretation of
bulges is complicated because internal dynamical effects in galaxies
might modify the kinematic properties of bulges.  For example, it has
been suggested by \citet{saha2012} that a
small bulge embedded in a bar can absorb angular momentum from the
bar, with a consequence that an initially non-rotating classical bulge
can transform into a cylindrically rotating triaxial object.  
\citet{saha2013} have also suggested that the appearance of B/P bulges
might be connected to the properties of dark matter halos.

\section{Stellar populations of B/P bulges}
\label{sec:6}

Stellar populations of bulges in barred and non-barred galaxies have
been compared using both absorption line-indices and applying stellar
population synthesis methods, but no clear conclusions are derived.
Using line-indices for 20 fairly face-on early-type barred galaxies
and comparing them with the non-barred galaxies by \citet{moorthy2006},
\citet{perez2011} found that
bulges in barred galaxies are more metal-rich and more
$\alpha$-enhanced than in non-barred galaxies. The
$\alpha$-enhancement is associated to rapid star formation event,
which is not expected if bulges formed via vertical buckling during
the formation and evolution of bars. Synthetic stellar population
methods has been applied for 62 barred and non-barred galaxies by
\citet{sanchez2014}, also for fairly face-on
galaxies. However, no difference in metallicity or age gradients
between barred and non-barred galaxies were found.  A sample of 32
edge-on galaxies was studied by \citet{jablonka2007},
and again no difference in the stellar populations of bulges
was found between barred and non-barred galaxies.

From our point of view critical questions are do the barred galaxies
in the above samples have B/P/barlens bulges or not, and what was
measured as a 'bulge'. For the first sample detailed morphological
classifications exist for 10 galaxies and it appears that even half of
those have barlenses in \citet{lauri2011}.  In the second sample 11
barred galaxies have detailed classifications, but none of them have
neither B/P nor barlens.  In \citet{perez2011} the bulges were taken to
be the central regions of the galaxies, which means also central
regions of barlenses. In most of the bulges studied by them star
forming nuclear rings and spiral arms were detected. In these fairly
face-on galaxies the star forming structures obviously had a strong
impact on the obtained stellar populations and metallicities, and do
not tell about the main stellar population of the B/P bulges.  It was
pointed out already by \citet{peletier2007} that composite bulges in
stellar populations indeed exist. And also, that due to dust in the
disk plane, the stellar populations and metallicities in the edge-on
and face-on views are expected to be different.

Clearly, understanding the different bulge components calls for
detailed studies of individual galaxies.  Based on the analysis of
four early-type galaxies \citet{sanchez2011} showed that most of the
stars in bulges are very old, (10 Gyr), as old as in the Milky Way
bulge. The same is true for bars, in which the stellar population ages
are closer to the bulges than to the disks outside the bars
\citep{perez2009}.  The stellar populations of the B/P bulges in 28
edge-on early-type disk galaxies (S0-Sb) have been studied by
\citet{williams2012} and \citet{williams2011}, and compared with the elliptical
galaxies. They looked at the properties both in the central regions,
covering the seeing-limited part of the boxy bulge, and in the main
body of the B/P structures. The central peaks were found to have
similar old stellar populations and high stellar velocity dispersion
as in elliptical galaxies. However, the main body of the B/P bulge
lacks a correlation between the metallicity gradient and $\sigma$,
which correlation appears in elliptical galaxies. Metallicity
gradients are easily produced in a monolithic collapse in the early
universe, and at some level also in violent galaxy mergers. But even
the non-barred galaxies in their study appeared to have stronger
metallicity gradients than the B/P bulges of the same galaxy mass.

In the literature it is often argued that the stellar populations of
bulges in S0-Sbc galaxies are similar to those of the elliptical
galaxies \citep{proctor2002,falco2006,macarthur2009}, which
similarity breaks only in the later type spirals
\citep[see][]{ganda2007}.  However, based on the analysis by
\citet{williams2012}, whether the bulges are similar to the
ellipticals or not, depends on what do we count as a bulge. If we mean
the central peaks in the radial flux distributions, then the answer is
that the bulges indeed are much like the elliptical galaxies, but if
we are talking about the main body of the B/Ps, then the stellar
populations are different from the elliptical galaxies.

An interesting example is NGC 357 \citep{lorenzo2012} for which galaxy
all critical diagnostics of B/P-structures have been made, including
the isophotal and stellar population analysis, kinematics, and
structural decompositions. They found that no single unambiguous
interpretation can be given for the bulge.  The galaxy has two bars,
which further complicates the interpretation. This example
demonstrates that based on the same analysis completely different
interpretations can be given for the bulge, depending on whether only
the central peak, with high rotation and $\sigma$ drop, is considered
as a bulge (i.e., a 'disky pseudobulge' in our notation), or the
larger region with high $\sigma$ is taken to be the bulge (i.e., the
classical bulge in our notation).  In their view, a problem in the
first interpretation is that the bulge has an old stellar population,
generally not accepted for a 'disky pseudobulges'.  If a classical
bulge is assumed then the problem is the nearly exponential surface
brightness profile. If the bulge is interpreted as a classical bulge
then there exists also a cool central disk inside that bulge.

But for the interpretation of this particular galaxy there exists also
a third possibility, namely that there is a boxy bulge and a central
cool disk embedded in that.  In this fairly face-on galaxy no B/P is
identified in the isophotal analysis, but the galaxy looks very much
like NGC 4643, in which galaxy a barlens has been recognized
\citep[see][]{lauri2014}.  The main bulge with the old stellar
population and a small S\'ersic index could simply corresponds to the
boxy bulge, which can also be dynamically fairly hot. The $\sigma$
drop could be associated to a central cool disk embedded in the boxy
bulge. It is worth noticing that the bulge has similar V-H color as
the rest of the galaxy, up to the outer radius of the bar
\citep{martini2003}, which fits into this interpretation. However, the
purpose of this paragraph is not to give the 'right interpretation',
but rather to demonstrate that not only detailed observations are
needed to study the bulges, but also the interpretation depends on the
current understanding of bulges.

\section{Summary and discussion}
\label{sec:7}

Nearly half of the highly inclined galaxies in the nearby universe are
found to have B/P/X-shape bulges. Barlenses, which appear in more
face-on galaxies, are likely to be physically the same
phenomenon. These structures appear in bright galaxies, in a mass
range near to the Milky Way mass. Also the other properties of these
bulges, including morphology (B/P/X-shape), kinematics (cylindrical
rotation), and stellar populations (old), are similar to those
observed in the Milky Way. Cool central disks are often embedded
inside the B/P/barlens bulges, in which case they are called as
composite bulges. Barred galaxies with composite bulges can contain
also dynamically hot classical bulges, but it is not yet clear to what
extent they, in the Milky Way mass galaxies, are really dynamically
distinct structure components, and to what extent stars wrapped into
the central regions of the galaxies during the formation and evolution
of bars.

A comparison of the observed properties of B/P/barlens bulges with the
simulation models have shown that they can indeed be explained as disk
structures, formed by buckling instabilities soon after the bars were
formed.  It is unlikely that any significant fraction of B/P bulges
were triggered by tidal effects.  Independent of the exact isophotal
shapes, these structures typically cover nearly half of the bar size,
but can be also smaller or larger than that. Also, recent structural
multi-component decompositions have shown that most of the bulge mass
in these galaxies might appear in the B/P/barlens bulges.  The
exceptions are the boxy structures appearing in the most massive
(M$>$10$^{11}$M$_\circ$) slowly rotating galaxies, which are often
elliptical galaxies (sometimes S0s) in classification. Exceptions are
also the 'thick boxy bulges', which might actually be manifestations
of thick disks in the otherwise almost bulgeless galaxies.

If we believe that most of the bulge mass in the Milky Way mass
galaxies indeed appears in the B/P/barlens bulges, it means that the
masses of the classical bulges must be very small in all Hubble types,
even in the early-type disk galaxies which are usually assumed to
contain most of the baryonic spheroidal mass.  If we are
unwilling to accept this conclusion, then it needs to be explained how
the observed morphological and kinematic properties of the
B/P/X-shape/barlens bulges are created in galaxies.  Why is this view
then not accepted as a paradigm in the astronomical community?
Actually, the argument that the B/P bulges form part of the bar, has
been accepted, but perhaps not the idea that such bar components could
contain most of the bulge mass in the nearby galaxies. One possible
reason for that is that the relative masses of bulges are generally
estimated from decompositions performed for fairly face-on
systems, in which galaxies the massive, round components are often
erroneously interpreted as classical bulges.

The explanations discussed for the formation of pseudobulges in this
review are related to the evolution of bars. A natural question is
then what makes the bulges in the non-barred galaxies?  If barred and
non-barred galaxies live in similar galaxy environments also the
accretion events should be similar, leading to bulge masses not too
different from each other. The relative masses of bulges in barred and
non-barred galaxies are compared for S$^4$G sample \citep{sheth2010}
of 2350 galaxies at 3.6$\mu$m by \citet{salo2015}. For galaxies
brighter than M*$>$10$^{10}$ $M_\circ$ larger bulge masses were found
for barred galaxies ($B/T\sim$0.15 and $B/T\sim$0.09,
respectively). Since in this study a photometric definition of a
'bulge' was used (excess flux above the disk in the surface brightness
profile), this difference most probably reflects the fact that in
barred galaxies part of the apparent bulge mass is associated to the
B/P/barlens bulge. There is some observational evidence that even in
the non-barred galaxies the bulges might have a B/P/barlens origin,
once the thin part of the bar has been dissolved
\citep[see][]{lauri2013}. The orbital analysis by \citet{patsis2002}
also predicts that peanuts may form even in galaxies without creating
any elongated, vertically thin bar components.  Naturally, there exists
also other ways of making pseudobulges in the non-barred galaxies, of
which minor mergers are one of the most prevalent
\citep[see][]{eliche2013}.

As a concluding mark we can say that in order to fully understand
bulges the same formative processes need to be valid both in the
edge-on and in face-on views. Also, it has become evident that bulges
are complex systems so that detailed studies of individual galaxies
are needed to separate the different bulge components. But even in
that case, the interpretation always reflects also the prevalent
theoretical understanding of a bulge, regardless of how sophisticated
diagnostics are used.
\vskip 0.3cm

% Heikki, miten nama tekstit saisi sisennettya ja pienemmalla fontilla?
%{\noindent \em \small Zu dem gebrauchten sowohl, wie zum dumpfen und stummen
%Vorrat der vollen Natur, den uns\"aglichen Summen,
%z\"ahle dich jubelnd hinzu und vernichte die Zahl.}
%\vskip 0.1cm

%{{\noindent  \em \small Summatta summautuu kaiken kaikkeuskiito,
%puhdas luonnon voima kun sit\"a luo.
%Lukemat unohtaen vain sen l\"oyd\"at - ja soit. }
%\vskip 0.15cm
%{\scriptsize Rainer Marie Rilke (Finnish translation by Liisa Enwald)}

\begin{acknowledgement}
We acknowledge the good and constructive comments of the referee, which helped
to improve this manuscript. We also acknowledge the
financial support from the Academy of Finland, and the DAGAL Marie Curie
Initial Training Network.  
.
%We acnowledge the referee Enrico Maria Corsini of constructive and insightful comments,
%which considerably improved our manuscript. 
\end{acknowledgement}

%\section{Conclusions}
%\label{sec:8}

%\bigskip
%\parindent=0pt
%{\bf References}
%\parindent=0pt

\end{document}